\documentclass[aps,prd,superscriptaddress,onecolumn,floatfix,nofootinbib,preprintnumbers]{revtex4}
\usepackage{amsmath,multirow}
\usepackage{graphicx,tabularx}
\usepackage{url}

\newcommand{\Meff}{{\ensuremath M}_\text{eff}}

\newcommand\one{\leavevmode\hbox{\small1\normalsize\kern-.33em1}}

\newcommand{\lag}{\mathcal{L}}

\newcommand{\met}{\slashchar{E}_T}

\newcommand{\tev}{{\ensuremath\rm TeV}}

\newcommand{\ifb}{{\ensuremath\rm fb^{-1}}}

\def\slashchar#1{\setbox0=\hbox{$#1$}           
   \dimen0=\wd0                                 
   \setbox1=\hbox{/} \dimen1=\wd1               
   \ifdim\dimen0>\dimen1                        
      \rlap{\hbox to \dimen0{\hfil/\hfil}}      
      #1                                        
   \else                                        
      \rlap{\hbox to \dimen1{\hfil$#1$\hfil}}   
      /                                         
   \fi}

\def\eg{{\sl e.g.} \,}
\def\ie{{\sl i.e.} \,}

\begin{document}

\title{Measuring Supersymmetry with Heavy Scalars}

\author{Emmanuel Turlay}
\affiliation{LAL, IN2P3/CNRS, Orsay, France}
\affiliation{NIKHEF, Amsterdam, Netherlands}

\author{Remi Lafaye} 
\affiliation{LAPP, Universit\'e de Savoie, IN2P3/CNRS, Annecy, France}

\author{Tilman Plehn}
\affiliation{Institut f\"ur Theoretische Physik, 
             Universit\"at Heidelberg, Germany}

\author{Michael Rauch}
\affiliation{Institut f\"ur Theoretische Physik, 
        Universit\"at Karlsruhe, KIT, Karlsruhe, Germany}

\author{Dirk Zerwas}
\affiliation{LAL, IN2P3/CNRS, Orsay, France}

\preprint{KA-TP-32-2010}
\preprint{SFB/CPP-10-93}

\begin{abstract}
  Supersymmetry with heavy scalars is a model where at the LHC we have
  to rely on rate measurements to determine the parameters of the
  underlying new physics. For this example we show how to properly
  combine rate measurements with kinematic endpoints, taking into
  account statistical, systematic and theory uncertainties. Provided
  we observe a sizeable number of events the LHC should be able to
  determine many model parameters with small enough error bars to for
  example test unification patterns.
\end{abstract}

\maketitle

\section{Setup}

Supersymmetry as a prime candidate for new physics at the TeV scale
should be discovered at the LHC even with moderate energy and
luminosity~\cite{review}. The challenge for complex TeV-scale
extensions of the Standard Model is to determine as many model
parameters as possible at the TeV
scale~\cite{sfitter,fittino,tev_scale} and extrapolate them to higher
energy scales. This way, we can study the underlying structures and
symmetries of an ultraviolet completion of our Standard Model --- up
to energy scales which might reach for example the scale of grand
unification~\cite{blair,sfitter_uni,Adam:2010uz,arthur}.

Most studies which focus on understanding new physics at and above the
TeV scale rely on a multitude of kinematic observables. In particular
at the LHC kinematic measurements are the most powerful, because they
can be extracted in the presence of large QCD and top-pair backgrounds
and are less prone to huge QCD corrections. Possible limits to such
strategies we have seen in LHC studies of supersymmetry with light
sleptons. There, the number of kinematic observables is drastically
reduced and the remaining kinematic features do not determine the
absolute new-physics mass scales well anymore~\cite{cambridge}. The
question then becomes how much information we can extract from fewer
and less robust observables, including production rates of
supersymmetric final states. Two aspects of such measurements mean
additional complications: first of all, we do not actually measure a
total signal cross section, but a matrix of production rates times
branching ratios in the presence of backgrounds and possibly relevant
kinematic cuts~\cite{Dreiner:2010gv}. Secondly, for such measurements the combination of
experimental and theory uncertainties becomes the crucial stumbling
block which determines if we can for example test gaugino mass
unification at the LHC or not. This situation is somewhat similar to
Higgs sector analyses at the
LHC~\cite{heavy_higgs,sfitter_higgs,Bock:2010nz}.\bigskip

In supersymmetry with decoupled scalars (DSS), all scalar partners are
decoupled from the relevant mass spectrum for the LHC
\begin{equation}
m_{\tilde{\ell}}=m_{\tilde{q}}=m_{H,H^\pm,A}\equiv m_S\geq 10^4~\text{GeV} 
\end{equation}
This scalar mass scale might be very
large~\cite{split_nima,split_gian} or simply decoupled from LHC
production above $\mathcal{O}(10~\tev)$~\cite{james}.  The observable
spectrum consists of the usual Standard Model fields, the gluino
$\tilde{g}$, the wino $\tilde{W}$, the bino $\tilde{B}$ and the
higgsino components $\tilde{H}_{u,d}$.  Omitting gauge-invariant
kinetic terms and non-renormalizable operators, the Lagrangian of the
low energy effective theory reads~\cite{split_gian,split_us,turlay_thesis}
\begin{eqnarray}
\lag & \supset & 
 m^2 H^\dagger H
-\frac{\lambda}{2}\left(H^\dagger H\right)^2 
- \Bigg[ \kappa_u\overline qu\epsilon H^*
       +\kappa_d\overline qdH
       +\kappa_e\overline\ell eH  \nonumber\\
     &&+\frac{1}{2} \left( M_3\hat g\hat g
                          +M_2\tilde{W}\tilde{W}
                          +M_1\tilde{B}\tilde{B} \right)
       +\mu\tilde{H}u^T\epsilon\tilde{H}d \nonumber\\
     &&+\frac{H^\dagger}{\sqrt{2}} \left(\hat g_u\sigma\tilde{W}
                                       +\hat g_u'\tilde{B} \right) \tilde{H}u
       +\frac{H^T\epsilon}{\sqrt{2}}\left(\hat g_d'\tilde{B}
                                         -\hat g_d\sigma\tilde{W} \right)\tilde{H}d
       +\text{h.c.}
\Bigg] \; ,
\end{eqnarray}
where $\sigma^a$ are the Pauli matrices ($\epsilon=i\sigma^2$). 
It includes one light Higgs doublet, tuned to have a small mass 
$m$ for $H=-\cos\beta\epsilon H^*_d+\sin\beta H_u.$ At the scale
$m_S$, the low-energy effective theory is matched to the full MSSM
\begin{alignat}{5}
\lambda(m_S) & = \frac{1}{4} \left[\frac{3}{5}g_1^2(m_S)+g_2^2(m_S)\right]
                             \cos^2 2\beta+\Delta_\text{th}, \notag \\
\kappa_u(m_S) & = y_u^*(m_S)\sin\beta, 
&\kappa_{d,e} (m_S) & = y_{d,e}^* (m_S) \cos\beta \notag \\
\hat g_u(m_S) & = g_2(m_S)\sin\beta, 
&\hat g_d(m_S) & = g_2(m_S)\cos\beta \notag \\
\hat g'_u(m_S) & = \sqrt{\frac{3}{5}}g_1(m_S)\sin\beta,
&\hat g'_d(m_S) & = \sqrt{\frac{3}{5}}g_1(m_S)\cos\beta
\label{eq:matching}
\end{alignat}
The four parameters $\hat{g}$ are the Yukawa couplings of the neutralinos and
charginos, which are modified with respect to their supersymmetric
values. Since it will be impossible to observe these deviations at the
LHC~\cite{split_us} we do not include them in our parameter extraction.
$\Delta_\text{th}$
are threshold corrections to the quartic Higgs coupling which change
the tree-level Higgs mass $m_h^2=2\lambda v$
\begin{equation}
\Delta_\text{th}= \frac{3y_t^4}{8\pi^2}
      \left[ \left(1-\frac{\frac{3}{5}g_1^2+g_2^2}{8y_t^2} \right)
              \frac{X_t^2}{m_S^2}
            -\frac{X_t^4}{12m_S^4}
      \right] ,
\end{equation}
with $X_t=A_t-\mu/\tan\beta$ in the same range as the gaugino and
higgsino masses.  The weak-scale parameter $\tan\beta$ only appears in
the boundary conditions and therefore is not a parameter of the
low-energy effective theory. It is interpreted as the fine-tuned angle
that rotates the two Higgs doublets into one heavy and one light mass
eigenstate~\cite{split_manuel}.\bigskip

New particles entering the renormalization group running at an
intermediate scale $m_S$ ($m_Z<m_S<M_\text{GUT}$) contribute
identically to the running of the three gauge couplings provided they
compose complete representations of the unification
group~\cite{sally}.  All sfermions in the MSSM form complete $SU(5)$
representations, so a possible gauge coupling unification scheme in
the MSSM is unchanged by heavy scalars. Experimentally establishing
such a pattern is one of the main long-term goals of the LHC.

In our analysis we construct the universal gaugino masses at the GUT
scale: $M_i(M_\text{GUT})=m_{1/2}$. They are then evolved down to the
scale $m_S$ based on one-loop renormalization group equations of the
MSSM~\cite{suspect,abdel}. The higgsino mass term $\mu$
is provided as an independent input parameter at the scale
$m_Z$. Below $m_S$ we integrate out all scalars and run the modified
renormalization group equations~\cite{split_gian} to the desired
scale.\bigskip

Keeping one of the supersymmetric Higgs bosons light requires a fine
tuning which for this study we accept without offering an
explanation. The Higgs mass matrix for the two Higgs scalars
\begin{equation}
\left(
\begin{array}{cc}
|\mu|^2+m_{H_u}^2 & b\\
b & |\mu|^2+m_{H_d}^2\\
\end{array}
\right)
\end{equation}
has eigenvalues $\langle m_H^2\rangle\pm\sqrt{\Delta^2+b^2}$ in terms
of $\langle m_H^2\rangle=(m_{H_u}^2+m_{H_d}^2)/2+|\mu|^2$ and
$\Delta=(m_{H_u}^2-m_{H_d}^2)/2$.  Requiring the light Higgs mass to
be of the order of the weak scale translates into
\begin{equation}
\label{eq:mhbound}
\sqrt{\Delta^2+b^2}-m^2_\text{ew}<\langle m_H^2\rangle<\sqrt{\Delta^2+b^2}.
\end{equation}
The $b$ term in the Lagrangian density breaks a Peccei-Quinn symmetry
and can therefore be kept small, as opposed to $\Delta$ and $\langle
m_H^2\rangle$ which should both be of order $m_S$. A light Higgs mass
means that $\langle m_H^2\rangle$ ranges around
$m_\text{ew}^2$. Hence, the fraction of the $(\langle
m_H^2\rangle,\Delta)$ space that satisfies Eq.(\ref{eq:mhbound})
corresponds to
\begin{equation}
\frac{V_\text{tuned}}{V_\text{total}} \sim
\frac{m_\text{ew}^2m_S^2}{m_S^4}\sim\frac{m_\text{ew}^2}{m_S^2}.
\end{equation}
\bigskip

Heavy scalars leave the chargino and neutralino spectrum untouched, so
the lightest neutralino should still be a good dark matter
agent~\cite{split_gian,split_dm}.  The measured density of dark matter
then imposes a constraint on $\mu$, which unlike in mSUGRA toy models
is not determined by electroweak symmetry breaking.  The WMAP
measurement $\Omega_\text{DM}h^2=0.111^{+0.006}_{-0.008}$~\cite{wmap}
can be reproduced in different parameter regions, which will be
represented by our choice of reference parameter points:
\begin{itemize}
\item[--] the `mixed region' with $M_1\approx\mu$ and a mixed
  higgsino-gaugino LSP. Here, $\tilde{\chi}_1^0\tilde{\chi}_1^0$ annihilation is
  enhanced for gauge and/or Higgs bosons or top quarks in the final
  state.
\item[--] the `pure higgsino' and `pure wino' regions where the LSP is
  almost mass degenerate with the $\tilde{\chi}^{\pm}_1$ and the
  $\tilde{\chi}^0_2$. This leads to an enhanced co-annihilation.  This region
  generally requires an LSP heavier than 1~TeV.
\item[--] the `Higgs pole' region in which the LSP is rather light,
  $m_{\chi^0_1}\approx \frac{1}{2}m_h$ and the annihilation proceeds
  via a resonant light Higgs.
\end{itemize}
\bigskip

If gluinos are lighter than squarks, they will mainly decay through
virtual squark exchange into quarks and
charginos/neutralinos~\cite{three_body}. For very heavy squarks
quantum corrections to the gluino decay processes can be significant,
because they are enhanced by the large logarithm
$m_{\tilde{g}}/m_S$~\cite{gambino}.  If the scalar mass scale is
larger than $10^6$~GeV, the gluino becomes sufficiently stable to form
$R$--hadrons~\cite{r_hadrons} which can be analyzed at the LHC without
major difficulties~\cite{split_us,stable}.  Those are subject to
cosmological constraints if they affect nucleosynthesis.  A gluino
with TeV-scale mass must have a lifetime shorter than 100~seconds to
avoid altering the abundances of deuterium and lithium-6. This sets an
upper limit of $m_S<10^9$~GeV~\cite{split_cosmo}. Therefore, in this
study we will focus on comparably short-lived gluinos.\bigskip

\begin{table}[t]
\begin{tabular}{l|rr}
& DSS1 & DSS2 \\
\hline
$m_S$             & \multicolumn{2}{c}{10~TeV}\\
$m_{1/2}(M_\text{GUT})$  & 132 GeV & 297~GeV\\
$\mu(m_Z)$         & 290 GeV   & 200~GeV\\
$\tan\beta(m_S)$  & \multicolumn{2}{c}{30}\\
$A_t(m_S)$        & \multicolumn{2}{c}{0}
\end{tabular}
\hspace*{30mm}
\begin{tabular}{c|rr}
& DSS1 & DSS2\\
\hline
$h$ & 129 & 129\\
$\tilde{g}$ & 438 & 880\\
\hline
$\tilde{\chi}^0_1$ & 60 & 125\\
$\tilde{\chi}^0_2$ & 117 & 185\\
$\tilde{\chi}^0_3$ & 296 & 206\\
$\tilde{\chi}^0_4$ & 310 & 317\\
\hline
$\tilde{\chi}^{\pm}_1$ & 117 & 175\\
$\tilde{\chi}^{\pm}_2$ & 313 & 317\\
\end{tabular}
\caption{Left: two parameter points chosen for this LHC study. Right:
  relevant mass spectrum from a modified version of {\tt
    SuSpect}~\cite{suspect}. All masses in GeV.}
\label{tab:DSS12param}
\end{table}

To quantitatively study supersymmetry with heavy scalars at the LHC we
define the two parameter points shown in Table~\ref{tab:DSS12param}.
They are in agreement with constraints from both dark matter
observations and collider searches.  A scalar mass scale of 10~TeV is
the lowest value which still qualifies as `decoupled' on LHC energy
scales. It limits the amount of fine-tuning in the Higgs sector to one
part in $10^4$ and gives us a short enough gluino life time to avoid
non-standard phenomenology or undesirable cosmological effects.  

The most important model parameters are $\mu$ and $m_{1/2}$. They
define the mass spectrum as well as the field content of the
neutralinos and charginos and most notably the LSP. Our two choices
span most of the LHC-relevant parameter space allowed by LEP and WMAP.
DSS1 lies in the Higgs pole region, so the LSP is mostly a bino with a
non-vanishing higgsino component. DSS2 lies in the `mixed region'.
The light LSP leads to an invisible Higgs branching ratio around
$1\%$. In DSS1, the gluino, the $\tilde{\chi}^0_2$ and $\tilde{\chi}^{\pm}_1$ are
fairly light and have correspondingly large LHC cross sections, as
shown in the Appendix.  For the heavier DSS2 spectrum this is no
longer the case.

The large value $\tan\beta=30$ avoids LEP limits on the light Higgs
mass. Its only impact on the low-energy theory is on $m_h$. The effect
of $A_t$ on $m_h$ is suppressed by at least $1/m_S^2$, so we can as
well set $A_t=0$.

\section{LHC observables}

Throughout our analysis we use leading-order and, where available,
next-to-leading-order Monte Carlo generators for signal and background
processes. The LO-generated samples are normalized to NLO
cross sections using $K$--factors.  For SUSY and $VV$ events we rely
on {\tt HERWIG}~\cite{herwig} including initial/final state radiation,
spin correlations in the decay of heavy states and angular
correlations between jets.  The $V+\text{jets}$ events we obtain from
{\tt ALPGEN}~\cite{alpgen} including matrix element and parton shower
matching~\cite{merging}.  The NLO normalization of these rates is
given by {\tt MCFM}~\cite{mcfm} or {\tt PROSPINO2}~\cite{prospino}.
Top pairs we simulate with {\tt MC@NLO}~\cite{mcnlo}, including its
NLO normalization. The pure QCD jets background we expect to be heavily
suppressed by cutting on, \eg missing energy.  Therefore, we leave it
at a leading-order {\tt PYTHIA} simulation~\cite{pythia}.
Detector and reconstruction effects we account for with a standard
general purpose detector simulation as described in Ref.~\cite{turlay_thesis}.\bigskip

Looking at the DSS signatures we rely on standard observables. They
are based on the reconstruction of isolated jets with $R=0.4$ and
leptons with $p_T>20$~GeV.  From the missing energy measurement we can
determine the effective mass $\Meff$ as a measure of the total
activity in the event.  It includes the four hardest jets and all
identified leptons.  For SUSY events, $\Meff$ scales with the mass of
the heavy particles produced and can be used to quantify the mass
scale of SUSY events~\cite{susy_inclusive,atlas_csc}.  The transverse
sphericity is defined as $S_T = 2\lambda_2/(\lambda_1+\lambda_2)$
where $\lambda_i$ are the eigenvalues of the $2\times2$ sphericity
tensor $S_{ij} = \sum_k p_{ki}p_{kj}$.  This tensor we compute using
all jets above $p_T= 20$~GeV and all leptons. SUSY events tend to be
relatively spherical ($S_T\sim1$) since the initial heavy particles
are usually produced approximately at rest and their cascade decays
emit particles in all directions. In the DSS1 parameter point the
sparticles are fairly light, resulting in a uniform distribution
between $S_T = 0.1$ and 0.6. For QCD jets or $V$+jets events $S_T$
peaks at zero. A cut on $S_T$ does not reduce the $t\bar{t}$
background as the distribution is very similar to the signal.\bigskip

The most frequent final states occurring for the DSS1/DSS2 parameter
point can be classified by the number of leptons: $0\ell$ (70\%/70\%),
$1\ell$ (20\%/23\%), $2\ell$ (5\%/5\%), and $3\ell$(3\%/1\%).  The number of
decay jets can vary from zero in the case of leptonically decaying
$\tilde{\chi}^{\pm}_1\tilde{\chi}^0_2$ pairs to eight in the case of gluino
pairs. This does not include jets from the underlying event or initial
and final state radiation which have to be carefully studied in
addition~\cite{susy_merging}.  The most common channel for both points
is the zero-lepton channel
$\tilde{g}\tilde{g}\rightarrow\tilde{\chi}^{\pm}_1\tilde{\chi}^{\pm}_1+4~\text{jets}\rightarrow8~\text{jets}+\met$. We
use it to estimate the discovery potential using standard
cuts~\cite{atlas_csc}: One hard jet with $p_T>100$~GeV and three over
50~GeV, no electrons or muons, an effective mass $\Meff>800$~GeV,
missing transverse energy $\met>100$~GeV with $\met>0.2\times\Meff$,
sphericity $S_T>0.2$ and finally a geometric separation between jets
and the missing energy vector of $\Delta\phi>0.2$ for the three
leading jets.  Both for jet mis-measurement and $b$ decays the $\met$
vector will be close the direction of one jet, so this cut reduces
fake missing energy from QCD. Note that for the various LHC
observables discussed below this basic set of cuts will be
modified.\bigskip

\begin{table}[t]
\begin{tabular}{l|r|rr}
& after cuts & \multicolumn{2}{c}{add'l $\Meff>$}\\
& & 800 GeV & 1 TeV \\
\hline
DSS1 & 12631 &4701 & \\
DSS2 &	145 & 112 & 97 \\
\hline
$t\bar{t}$ & 5161 &	274 & 62\\
QCD jets  & 848 & 15&\\
$W$+jets &	769 &	195&\\
$Z$+jets & 422 & 128&\\
$WZ$ & 20& 4&\\
$WW$ & 12& 3&\\
$ZZ$ & 2& 2&\\ \hline
total SM & 7234 & 621 & 62 \\ \hline
DSS1 significance & & 18&\\
DSS2 significance & & 0.8&6\\
\end{tabular}
\caption{Number of events remaining for each process after the cuts
  listed in the text. They are normalized to NLO cross sections for an
  energy of 14~TeV and an integrated luminosity of $1~\ifb$.}
\label{tab:yield}
\end{table}

Table~\ref{tab:yield} shows the number of events remaining after cuts.
After all but the $\Meff$ cut $t\bar{t}$ is the dominant background,
but there are also significant contributions from $V$+jets. Finally,
$\Meff > 800$~GeV reduces the background to below the level of the
signal for DSS1.

The systematic uncertainties on the number of background events for
$1~\ifb$ we take to be 50\% for QCD multi-jets and 20\% for
$t\bar{t}$ and $V$+jets, $WW$, $WZ$ and $ZZ$. This corresponds to a
combination of data-driven and Monte Carlo methods~\cite{atlas_csc}.

A discovery of new physics can then be claimed if the number of
observed events exceeds 25 and the significance is larger than
five. The significance of the observation of DSS1 with an integrated
luminosity of $1~\ifb$ is 18, so this parameter point will be
discovered at the LHC within one year of data taking at low
luminosity. For DSS2, the significance is low and, for this set of
cuts, does not increase with statistics. However, if we require
$\Meff>1$~TeV the significance for $1~\ifb$ increases to 6.

\subsubsection*{Higgs mass}

With decoupled scalars the light Higgs scalar is essentially
equivalent to its Standard Model counter part at 129~GeV.  Its mass
depends on $M_S$ and on $\tan\beta$. The $M_S$ dependence arises from
the running of $\lambda$ from $m_S$ to the weak scale, while
$\tan\beta$ appears as $\cos^22\beta$ in the matching. This can impact
the numerical value of $m_h$ by up to 20~GeV and should --- like
usually in supersymmetric Higgs studies --- allow us to determine
$\tan\beta$ at the LHC.

The total next-to-leading order production cross section is 39~pb. It
is computed by {\tt HIGLU} for gluon fusion~\cite{spirix}, {\tt VV2HF}
for weak boson fusion and {\tt V2HV} for the production in association
with a vector boson. This number includes NLO QCD corrections, and for
the first one also the NLO electroweak
contributions~\cite{Actis:2008ts}. The
dominant decays are into $b\bar{b}$ (53\%), $WW^*$ (29\%), $\tau\tau$
(5\%), and $ZZ$ (4\%)~\cite{hdecay,s_hit}.  In addition, the branching
ratio into photons reaches its maximum of 0.2\%, allowing for a precise
mass measurement.

Higgs production through supersymmetric cascades only occurs in
$\tilde{\chi}^0_{3,4}$ and $\tilde{\chi}^{\pm}_2$ decays. The corresponding production
rates are small, so their contribution to the total Higgs production
is of the order of 100~fb, \ie negligible compared to SM channels.

Systematic uncertainties on the Higgs mass measurement arise from the
electromagnetic energy scale.  The calibration of the photon energy
scale will be derived from $Z\rightarrow ee$ events and
$Z\rightarrow\mu\mu\gamma$ events, with an expected accuracy of 0.1\%.
The statistical uncertainty for an integrated luminosity $100~\ifb$
should also range around 0.1\%~\cite{atlas_csc}. The theory
uncertainty due to higher-order corrections to $m_h$ should not exceed
the very conservative limit of 4\%~\cite{higgsmass}.

\subsubsection*{Di-lepton endpoint}

Kinematic endpoints are usually the main ingredients to supersymmetric
parameter analyses, due to their small experimental and theory
errors~\cite{edges,gluino_mass,edges_nlo}. Perfect triangular
di-lepton edges in cascade decays occur in successive two-body decays,
like $\tilde{\chi}^0_2 \to \tilde{\ell} \ell \to \tilde{\chi}^0_1 \ell \ell$.  
Such a measurement directly
constrains the gaugino mass parameters $M_1$ and $M_2$ as well as the
higgsino mass parameter $\mu$.  In contrast, decays via on-shell $Z$
bosons only give us the $Z$ mass peak with little information on the
supersymmetric masses in the decay. For our parameter choices,
$\tilde{\chi}^0_3$ (DSS1) or both $\tilde{\chi}^0_{3,4}$ (DSS2) as well as the
charginos in DSS2 decay through such an on-shell gauge
boson~\cite{s_hit,sdecay}.

However, we can also apply endpoint techniques to three-body decays:
$\tilde{\chi}^0_2\rightarrow\tilde{\chi}^0_1\ell\ell$ in both points and
$\tilde{\chi}^0_3\rightarrow\tilde{\chi}^0_1\ell\ell$ in DSS2 lead to two leptons with
opposite signs and same flavor (OSSF), as listed in the Appendix.  On
the production side, the $\tilde{\chi}^{\pm}_1\tilde{\chi}^0_2$ channel has little
background due to the small number of jets in the final state. To
increase the total rate we also include $\tilde{\chi}^0_2$ production from
gluino decays. The total available cross section leading to this decay
becomes $\sigma(\tilde{\chi}^0_2\rightarrow\tilde{\chi}^0_1\ell\ell)\approx 3.5$~pb
and 93~fb in DSS1 or DSS2 and
$\sigma(\tilde{\chi}^0_3\rightarrow\tilde{\chi}^0_1\ell\ell)\approx 75$~fb in
DSS2.\bigskip

In addition to the staggered jet cuts ($p_{T,j} > 100,50,...$~GeV) we
now require at least two OSSF electrons or muons with $p_T>20$~GeV,
$|\eta|<2.5$, and $m_{\ell\ell}<m_{\chi^0_2}-m_{\chi^0_1} +
10$~GeV. Since the true value of the endpoint is a priory unknown,
this choice implies that the edge has already been observed.  To
remove combinatorial as well as top backgrounds we apply flavor
subtraction. Many backgrounds cancel in the combination
\begin{equation}
\frac{N(e^+e^-)}{\beta} +\beta N(\mu^+\mu^-) -N(e^\pm\mu^\mp)
\end{equation}
where $\beta$ is an efficiency correction factor equal to the ratio of
the electron and muon reconstruction efficiencies.\bigskip

In the DSS2 parameter point, the lowest endpoint corresponds to the mass
splitting $m_{\chi^0_{2,1}}$ while the second one corresponds to
$m_{\chi^0_{3,1}}$. We fit its distribution with a superposition of
three components, two modeling the decay kinematics plus a
Breit-Wigner $Z$ line shape.  Table~\ref{tab:Mllfit} compares the
results with the theoretical values.  The extracted values are in good
agreement with the input values for DSS1 and in reasonable agreement
for DSS2.  The statistical errors on the mass differences we can
extract from the fit to the reference function. Systematic
uncertainties are dominated by the lepton energy scale (0.1\%), while
theory errors due to unknown higher-order contributions are expected
to range around a percent.

There might occur doubts if the second edge giving
$m_{\chi^0_{3,1}}$ is actually visible, so we check that the curve
between the $m_{\chi^0_{2,1}}$ edge and the onset of the $Z$ peak
indeed lies $5\sigma$ above the background-only prediction.

\begin{table}[t]
\begin{tabular}{cc|c|c|c}
& & theory & fit value & statistical error\\
\hline
\multicolumn{1}{c|}{DSS1} & $m_{\chi^0_2}-m_{\chi^0_1}$ & 55.1 & 55.2&
$\pm0.6$ / 10 fb$^{-1}$\\
\hline
\multicolumn{1}{c|}{DSS2} & $m_{\chi^0_2}-m_{\chi^0_1}$ & 60.7 & 60.2&
$\pm2$ / 100 fb$^{-1}$\\
\multicolumn{1}{c|}{} & $m_{\chi^0_3}-m_{\chi^0_1}$ & 81.9 & 79.0&
$\pm3$ / 100 fb$^{-1}$\\
\end{tabular}
\caption{Results of the fit to the invariant mass distribution. All
  values given in GeV.}
\label{tab:Mllfit}
\end{table}

Assigning these measured values to sparticle mass differences
necessitates a few assumptions. In DSS1, lepton pairs are
quite frequent with respect to the overall SUSY production. This
suggests that the neutralino triggering this decay is somewhat
light. In addition, a decay through a 296 GeV~$\tilde{\chi}^0_3$ is rather
unlikely, and in such a case additional structure would be seen. In
DSS2, with two endpoints and a $Z$ peak, the interpretation is
more complicated. In addition to the assumption that
the endpoints arise from $\tilde{\chi}^0_{2,3}$ decays  we have to
assume that $\tilde{\chi}^0_3$ decays preferably to $\tilde{\chi}^0_1$. Otherwise, the
largest endpoint could correspond to $m_{\chi^0_{3,2}}$ and the
$Z$ peak to the decay $\tilde{\chi}^0_3\rightarrow\tilde{\chi}^0_1Z$.

\subsubsection*{Di-jet endpoint}

Unfortunately, the technique described above is only applicable to
decays involving $b$ jets, but not light-flavor jets.  In DSS1, 1.7\%
of gluinos decay to the LSP with two bottom quarks.  We select these
events by requiring four jets with $E_T>50$~GeV, no leptons and
$\met>100$~GeV. For two $b$-tagged jets we compute the invariant mass
$m_{bb}$. The background is dominated by $t\bar{t}$ events, as well as
combinations due to decays other than $\tilde{g}\rightarrow\tilde{\chi}^0_1
b\bar{b}$.  The fit output from $m_{bb}$ we compare to the theoretical
values: for $10~\ifb$ the fit value of 380.6~GeV corresponds to the
input of 383.0~GeV within the statistical error of 5.2~GeV.  This
measurement is the basis for the determination of $M_3$ from the
gluino mass~\cite{gluino_mass}.

\subsubsection*{Tri-lepton cross section}

\begin{table}[t]
\begin{tabular}{c|r|c|r}
DSS1 & $\sigma$ & BR
&$\sigma(3\ell)$\\
\hline
direct & 11.7 pb & 1.54\% &180 fb\\
via $\tilde{g}\tilde{g}$ & 10.4 pb & &160 fb\\
\hline
\multicolumn{2}{c|}{} & total & 340 fb\\
\end{tabular}
\hspace*{2cm}
\begin{tabular}{c|r|c|r}
DSS2 & $\sigma$ &  BR &$\sigma(3\ell)$\\
\hline
direct & 1390 fb & 1.54\% & 21.4 fb\\
via $\tilde{g}\tilde{g}$ & 166 fb & & 2.6 fb\\
\hline
\multicolumn{2}{c|}{} & total & 24 fb\\
\end{tabular}
\caption{Cross sections contributing to the tri-lepton signal for the LHC
 running at 14~TeV.}
\label{tab:3l}
\end{table}

In contrast to the usual and more optimistic
scenarios~\cite{sps1a,gluino_mass,sfitter}, decoupling all scalars at
the LHC implies that we will not have enough kinematic information to
extract masses and model parameters of the underlying new-physics
model. Therefore, we need to rely on cross section measurements, in
spite of their larger experimental and theory errors. The main purpose
of this analysis is to show how such rate measurements can indeed be
used as input to new physics measurements.

As usually, signatures involving leptons have lower LHC backgrounds
and increase the precision of the measurement. For heavy
supersymmetric scalars charginos and neutralinos will give different
final states with numerous isolated leptons. The tri-lepton final
state allows for background rejection by requiring two OSSF
leptons. It arises from chargino and neutralino production with
subsequent decays $\tilde{\chi}^0_2\rightarrow\tilde{\chi}^0_1\ell\ell$ and
$\tilde{\chi}^{\pm}_1\rightarrow\tilde{\chi}^0_1\ell\nu$.  In Table~\ref{tab:3l} we
show the composition of the tri-lepton signal in DSS1. Pairs of
$\tilde{\chi}^{\pm}_1\tilde{\chi}^0_2$ are produced directly as well as in gluino
decays.\bigskip

We select these events requiring at least one OSSF pair and exactly
three leptons. In case of direct production, two LSPs will be emitted
essentially back-to-back, hence canceling the missing transverse
energy, so we lower our cut to $\met>50$~GeV. An optional jet veto
above $p_T \sim 20$~GeV (dependent on detailed QCD studies) can be
applied in order to select events from direct $\tilde{\chi}^{\pm}_1\tilde{\chi}^0_2$
production rather than $\tilde{g}$ pair decays. However, its effect on
signal and background rates is hard to predict, so the results should
be taken with a grain of salt.  To reject $Z$ decays we veto
$m_{\ell\ell}=81.2...102.2$~GeV. For our signal, we expect
$m_{\ell\ell} \lesssim 56$~GeV due to the $\tilde{\chi}^0_{2,1}$ mass
splitting.

In Table~\ref{tab:yield3l} we present the number of events after
cuts.  Only $t\bar{t}$ and $WW/WZ/ZZ$ are significant, where the
latter is already partly removed by the $m_{\ell\ell}$ cut while the
former can be removed by a jet veto.\bigskip

An as precise as possible extraction of the number of tri-lepton
signal events relies on our knowledge of the backgrounds and a
complete understanding of detector effects, luminosity, parton
distributions, and finally cut efficiencies.  The systematic
uncertainty on $\sigma(\text{SUSY} \to 3 \ell)$ is certainly bounded
from below by the knowledge of the luminosity $\mathcal L$. \ie of the
order of 5\%~\cite{atlas_lumi}. In order to take into account
additional systematic errors we consider cases of 5\%, 10\% and
20\%. The theory uncertainty due to QCD effects we estimate to be of
the order of 12\%~\cite{prospino}. 

\begin{table}[b]
\begin{tabular}{l|r|r}
& after cuts & add'l jet veto\\
\hline
DSS1 & 681& 43  \\
DSS2 &  87& 4\\
\hline
$t\bar{t}$ &1,106 &59 \\
QCD        &0 &0 \\
$W/Z+\text{jets}$ & 14&0 \\
$WW/WZ/ZZ$        &235 &73 \\
\end{tabular}
\caption{\label{tab:yield3l}Number of tri-lepton events remaining
  after the cuts discussed in the text (assuming $10~\ifb$).}
\end{table}

\subsubsection*{Gluino pair cross section}

The gluino pair channel constitutes a large fraction of the signal
events in both parameter points (77\% in DSS1 and 22\% in
DSS2). Different strategies should allow us to select only
$\tilde{g}\tilde{g}$ events.  However, they will be very model
dependent, as they require at least a guess of the gluino decays
branching fractions. We can for instance take advantage of the very
short zero-lepton cascade
$\tilde{g}\tilde{g}\rightarrow\tilde{\chi}^0_1\tilde{\chi}^0_1q\bar{q}q\bar{q}$
to remove SUSY background from chargino and neutralino channels.
Again, the systematic uncertainty is bounded from below by the
knowledge of the luminosity, so we consider systematic errors of 5\%,
10\% and 20\%. Theory uncertainties from higher order contributions
can be 30\%, estimated from a scale variation by a factor $1/4 ... 4$
at next-to-leading order~\cite{prospino}. It is meant to be
conservative to also accommodate additional sources of uncertainties,
like parton densities and the strong coupling.  It can be reduced once
we systematically include higher-order QCD corrections to the
production of heavy particles~\cite{nnlo}. This choice of theory
errors will give us a conservative estimate if including rate
information at the LHC should become a part of supersymmetric
parameter analyses.

\subsubsection*{On-shell vs off-shell $Z$ bosons}

If the mass difference between two neutralinos or two charginos is
larger than $m_Z$, the invariant mass distribution of its decay
products will exhibit a very sharp peak. In contrast, if the mass
difference between the two sparticles is too small, the $m_{\ell\ell}$
distribution will show a triangular shape with a sharp endpoint below
$m_Z$.  This effect can provide valuable information about the
neutralino and chargino masses as well as their couplings to the $Z$
boson.

The measurement of the $\sigma(\text{SUSY}\rightarrow Z)$ rate
requires a good knowledge of the luminosity, lepton efficiencies and
background rates. Some of these source of systematic uncertainties
cancel from the ratio
\begin{equation}
\label{eq:RZ}
R_Z=\frac{N\left(m_{\ell\ell}>\text{endpoint}\right)}
         {N\left(m_{\ell\ell}<\text{endpoint}\right)} \; .
\end{equation}
For our two reference points we find $R_Z < 0.004 \sim 0$~(DSS1) and
0.196~(DSS2), respectively.  Systematic errors arise if the $p_T$
spectra of leptons from the $Z$ peak and from the triangular shape are
different or if the identification efficiencies for these two spectra
are different. However, electron identification efficiencies are
reasonably flat for $p_T>25$~GeV. Leptons from the $Z$ peak will have
transverse momenta around 45~GeV, while those under the triangle curve
we cut to $p_T>20$~GeV.  We therefore include an overall 1\%
systematic error to mainly account for lepton identification
uncertainties.  The theory uncertainty due to the prediction of
branching ratios should be of the order of 1\%~\cite{s_hit,sdecay}.

\section{Parameter determination}

Once new physics (\eg supersymmetry) will be discovered at the LHC, we
have to turn our focus towards understanding the corresponding
signatures.  On the one hand, we will have to test different types of
TeV-scale models with the available data, while on the other hand we
have to consistently determine the parameters of the underlying theory
for each of these model hypotheses. This might well include combining
LHC observables with other measurements such as the relic density of
dark matter, the magnetic moment of the muon, or flavor physics. Note
that a consistent approach does not allow for the replacement of some
measurements by top-down predictions in someone's favorite
model. Instead, we need to see how far we can get, for example with
the limited set of observables described in the last section.\bigskip

SFitter~\cite{sfitter,sfitter_higgs} is designed to map up to
20-dimensional highly complex parameter spaces onto a large set of
observables of varying quality, which can be highly correlated.  It
can be used to estimate the reach in terms of a given model for any
experiment, but also to realize a proper bottom-up approach to
determine the parameters of a fundamental theory.

\begin{table}[t]
    \begin{tabular}{l|l|c|l|l|c|r}
\multicolumn{3}{c|}{} & \multicolumn{4}{c}{uncertainties (\%)}\\
      \cline{4-7}
\multicolumn{1}{c|}{}      &\multicolumn{2}{c|}{observables} & stat. & \multicolumn{2}{c|}{systematic} & th.\\
      \cline{5-6}
\multicolumn{1}{c|}{}&\multicolumn{2}{c|}{}  & &value & source & \\
      \hline
       DSS1 &$m_h$ & 129 GeV & 0.1 &0.1 & energy scale &   4\\
      &$m_{\chi_2^0}-m_{\chi_1^0}$ & 55.2 GeV & 1 &0.1 & energy
      scale &  1\\
      &$m_{\tilde g}-m_{\chi_1^0}$ & 382.8 GeV & 1.5 &1 & energy scale & 
      1\\
      &$\sigma(3\ell)$ & 340 fb & 2 &$>5$ & luminosity &   12\\
      &$R_Z$ & $<0.004$ & 0.01 &1 & lepton id. & 1\\
      &$\sigma(\tilde{g}\tilde{g})$ & 62.8 pb & 0.1&$>5$ & luminosity &   30\\
      \hline
      DSS2 &$m_h$ & 129 GeV & 0.1 &0.1 & energy scale &   4\\
      &$m_{\chi_2^0}-m_{\chi_1^0}$ & 60.7 GeV &  3.3
      &0.1 & energy scale &  1\\
      &$m_{\chi_3^0}-m_{\chi_1^0}$ & 81.9 GeV &  3.7
      &0.1 & energy scale &  1\\
      &$\sigma(3\ell)$ & 23 fb & 14 &$>5$ & luminosity &   12\\
      &$R_Z$ & 0.57 & 0.7 &1 & lepton id. & 1\\
      &$\sigma(\tilde g\tilde g)$ & 1067 fb &  3&$>5$ & luminosity &
      30\\
    \end{tabular}
\hspace*{20mm}
\begin{tabular}{c|c|r|r|r|r|r|r|r|r}
\multicolumn{2}{c|}{} & 
\multicolumn{2}{c}{\sc EXP}   &
\multicolumn{2}{|c|}{\sc SYST} &
\multicolumn{2}{c|}{\sc TH}     &
\multicolumn{2}{c}{\sc FULL}\\
\cline{3-10}
\multicolumn{2}{c|}{} & 
\multicolumn{1}{c|}{$\Delta$}&\multicolumn{1}{c|}{\%} & \multicolumn{1}{c}{$\Delta$} & 
\multicolumn{1}{|c|}{\%} &\multicolumn{1}{c|}{$\Delta$} & \multicolumn{1}{c|}{\%} &
\multicolumn{1}{c|}{$\Delta$} & \multicolumn{1}{c}{\%} \\
\hline
DSS1 &$M_1$       & 2.8 & 2.1 & 1.1 & 0.8 & 4.8 & 3.6 & 5.5 & 4.1 \\
     &$M_2$       & 3.0 & 2.2 & 1.5 & 1.1 & 2.5 & 1.9 & 3.8 & 2.9 \\
     &$M_3$       & 0.9 & 0.7 & 0.7 & 0.5 & 2.2 & 1.7 & 2.4 & 1.8 \\
     &$\mu$       & 53  & 18  & 26  & 8.9 & 37  & 13  & 47 & 16  \\
     &$\tan\beta$ & 21  & 66  & 14  & 46  & 12  & 40  & 14 & 47  \\
\hline
DSS2 & $M_1$       & 14.8 & 5.0 & 2.1 & 0.7 & 9.4  & 3.1 & 14.6 & 4.9 \\
     & $M_2$       & 6.8  & 2.3 & 0.9 & 0.3 & 4.7  & 1.5 & 7.1  & 2.4 \\
     & $M_3$       & 2.3  & 0.8 & 0.7 & 0.2 & 12.3 &  4  & 15.6 & 5.3 \\
     & $\mu$       & 7.2  & 3.6 & 1.0 & 0.5 & 4.7  & 2.4 & 7.0  & 3.5 \\
     & $\tan\beta$ & 20   & 67  & 1.6 & 5.3 & 16   & 53  & 20.6 & 69  \\
\end{tabular}
\caption{Left: summary of available collider observables
  in DSS1 and DSS2. Statistical errors are quoted for an integrated
  luminosity of $100~\ifb$.  Right: absolute and
  relative errors on the determination of the underlying model
  parameters for three fitting strategies described in the text.}
\label{tab:DSSobs}
\end{table}

The determination of the parameters then proceeds in two steps. First,
we maximize the exclusive log-likelihood using a weighted Markov
chain~\cite{sfitter} to identify the best-fitting point in parameter
space.  The starting point of this Markov chain is arbitrary, and we
repeat the search several times to ensure our procedure converges
well. This minimum then serves as starting point for a {\tt MINOS} hill-climbing
minimization to improve the resolution and to estimate the
errors.\bigskip

In models with decoupled scalars we can use the different LHC
observables discussed above to determine the parameters of the model.
Table~\ref{tab:DSSobs} summarizes them along with their expected
uncertainties for both parameter points.  For the mass differences we
use the result of detailed experimental analyses, while for the
rate-related observables we rely on the theoretical central value,
lacking the complete experimental analysis.

By definition, no information on the squark and slepton sector is
available except for its explicit absence. Consequently, we fix $m_S$
and $A_t$ to large (nominal) values. The three gaugino mass parameters
we fit independently, to allow for a bottom-up experimental test of
gaugino mass unification. Technically, we know that for scalar masses
a consistent bottom-up approach does not reproduce the usual top-down
results, which means we would have to evolve all parameters strictly
from the weak scale to the high scale~\cite{Adam:2010uz}. For gauginos
the differences between the two methods are not as large, so for
illustration purposes we use a top-down running for the
renormalization group equations.\bigskip

Four scenarios illustrate well the precision we can reach on the
determination of supersymmetric model parameters with decoupled
scalars. We use about 1000 toy experiments for each scenario and
each benchmark point. The toy experiments are generated by smearing
the observables according to the expected experimental and/or
theoretical errors, depending on the scenarios. Correlations
among the measurements are taken into account separately
for the energy scale of leptons, jets and the luminosity measurements.
For each toy experiment the best-fit parameter set is determined.
From the distribution of the best-fit parameter we read off the error as
the RMS (Root-Mean-Square) of the distribution. Table~\ref{tab:DSSobs}
displays the resulting uncertainties for the four scenarios:
\begin{itemize}
\item[--] EXPerimental errors. In order to evaluate
  the impact of the pure experimental uncertainties on the determination of
  the DSS parameters for a given luminosity of $100~\ifb$, we take into
  account the statistical and systematic errors, but we do not include theory errors.

  We see that with roughly a $50\%$ error on $\tan \beta$ we can
  hardly determine this parameter in both reference points. Because
  all heavy Higgs bosons are decoupled our only leverage
  is the light Higgs mass which depends on
  several parameters, including $m_S$ in the (s)top sector. The better
  way to study the Higgs sector, possibly including $\tan\beta$, would
  be a dedicated Higgs analysis at the LHC~\cite{sfitter_higgs,Bock:2010nz}. 

  The second Higgs-sector parameter $\mu$ is determined to better than
  1\% in DSS1 and 4\% in DSS2. In the former, this is due to large
  production rates for neutralinos and charginos. In the latter, the
  higgsino fractions are well spread over all neutralinos, so
  neutralino mass differences include this information. Thanks to very
  small statistical uncertainties on all mass differences in DSS1,
  also the gaugino mass parameters are determined within 2\%. This is
  not the case in DSS2, implying a deterioration to $\sim10\%$. As
  discussed above, the observability of the second neutralino edge is
  arguable. We perform the DSS2 fit with and without this
  observable and find identical results.

\item[--] SYSTematic errors. Here we assume that a very large number
  of events has been gathered at the LHC ($\geq 300~\ifb$) and that
  statistical uncertainties are negligible. By that time, we might
  assume that theoretical predictions of $m_h$ and the different
  cross sections will have rendered the theory error negligible as well. This
  idealized scenario is useful to measure the impact of systematic
  errors and the ultimate precision of the parameter determination.

  In Table~\ref{tab:DSSobs} we observe that with only systematic
  errors all parameters are determined within 1\%, except for
  $\tan\beta$ which still suffers from an invisible extended Higgs
  sector. Systematic errors on cross section measurements we vary
  between of 5 and 20\%. In DSS1, this hardly affects the parameter
  determination as all model parameters are already well constrained
  by mass measurements. In DSS2, a 20\% systematic error on
  $\sigma(\tilde{g}\tilde{g})$ doubles the uncertainty on $M_3$ as
  compared to the case where we are dominated by luminosity
  measurement (5\% systematic error). In contrast, a 20\% systematic
  error on $\sigma(3\ell)$ again does not affect the parameter
  determination, because the weak masses are constrained by kinematics
  measurements.

  Compared to the first case including statistical and systematic
  uncertainties we observe a gain of at least a factor~4 in the errors
  for DSS2. This shows that for the integrated luminosity of
  $100~\ifb$ assumed in the scenario EXP the statistical error still
  dominates.

\item[--] THeory errors. Again, we assume very large statistics but
  include theory errors, thus taking into account theoretical and
  systematic errors. This scenario gives us a flavor of what is
  achievable after at least five years of operation of the LHC at full
  luminosity.

  With a 4\% theory uncertainty on $m_h$, $\tan\beta$ is now
  practically undetermined. The interplay of neutralino mass
  splittings, $R_Z$ and the tri-lepton cross section provides enough
  constraints on the neutralino and chargino sector to allow for a
  determination of $M_1$, $M_2$ and $\mu$ to better than 4\%.  In DSS1, $M_3$
  is very well determined by the mass splitting
  $m_{\tilde{g}}-m_{\chi^0_1}$.  However, in DSS2 the only available
  handle on $M_3$ comes from the gluino pair cross section which
  suffers from a large theoretical uncertainty which we set conservatively to 30\%. 
  As the production cross section
  strongly decreases as function of $M_3$, the distribution of the 
  toy experiments resulting from the symmetric theoretical error is highly asymmetric
  with a long tail to large values of $M_3$ as expected. In spite of the 
  large theoretical error on the cross section prediction, a $M_3$ measurement
  significantly better than at the 10\% level is feasible.

\item[--] FULL errors. In this scenario we combine the experimental errors (EXP)
with the theoretical errors (TH) on the observables for $100~\ifb$.
In DSS1, the errors on the gaugino masses are slightly larger than the errors
for the TH scenario.
In DSS2 the errors on $M_1$ and $M_2$ are essentially the same 
as the EXP errors, the corresponding mass difference measurements are
dominated by the statistical error. For $M_3$ the experimental
errors are fairly small compared to the theoretical error. Nevertheless,
the determination remains at the level of several percent due to the steep 
descent of the cross sections as a function of the gluino mass.
Including further theoretical developments is necessary to increase the precision 
of the parameter determination by reducing the theoretical error~\cite{nnlo}.

\end{itemize}

\section{Outlook}

Supersymmetry with heavy scalars is a variation of the MSSM which more
naturally accommodates for example flavor constraints, at the expense
of the solution of the hierarchy problem. For the LHC, it is
irrelevant as which energy scale the scalars reside, as long as they
are at least of the order of $10^4$~GeV.

Such a model is a serious challenge to any kind of supersymmetric
parameter analysis, because it severely reduces the number of LHC
observables, in particular from cascade decay kinematics. Instead, we
need to rely for example on rate measurements, including their complex
systematic and theory error structure. Of our two parameter points the
first one should be discoverable within a year of data-taking at the
LHC, due to very large SUSY cross sections. The second point has
lower rates but should be discovered within a few years. For both
scenarios we establish a set of observables, including statistical,
systematic, and theory uncertainties.\bigskip

For a model with heavy scalars we have shown that a global fit of the
model parameters to the experimental observables is still able to
determine the correct central values and corresponding errors for all
parameters.  The weakly interacting sector ($M_1$, $M_2$ and $\mu$)
can be fairly well measured at the LHC, with conservative accuracies of
the order of a few percent after including all error
sources. Most notably, this includes all uncertainties related to rate
measurements at the LHC. The Higgs sector suffers from the fact that
we will only have one observable at hand, namely the light Higgs
mass. For such a situation we will have to resort to a dedicated Higgs
sector analysis at the LHC~\cite{sfitter_higgs,Bock:2010nz}.

Even for optimistic LHC luminosities and energies the impact of
systematic uncertainties is likely not dominant, even though an
estimate of these errors prior to a full-fledged analysis on real data
should be taken with a grain of salt. For all theory uncertainties we
have relied on particularly conservative estimates, which means that
the positive outcome of our study is generally dependable.

\acknowledgments All of us thank Giacomo Polesello for numerous
discussions. Dirk Zerwas would like to thank Heidelberg University for
the hospitality during his frequent visits. Part of this work was
developed in the framework of the CNRS GDR Terascale. MR acknowledges
support by the Deutsche Forschungsgemeinschaft via the
Sonderforschungsbereich/Transregio SFB/TR-9 `Computational Particle
Physics' and the Initiative and Networking Fund of the Helmholtz
Association, contract HA-101 (Physics at the Terascale).

\appendix

\section{Production rates and branching ratios}

\begin{table}[t]
\begin{tabular}{l|rrrr||l|rrrr}
 & \multicolumn{2}{c}{DSS1} &\multicolumn{2}{c||}{DSS2} &
 & \multicolumn{2}{c}{DSS1} &\multicolumn{2}{c}{DSS2}\\
\hline
$\tilde{g}\tilde{g}$ & \multicolumn{2}{c}{62.8 pb} & \multicolumn{2}{c||}{954 fb} &
$\tilde{\chi}^0_1\tilde{g}$ & 71 fb&  & 0.01 fb &\\
\cline{1-5}
$\tilde{\chi}^{\pm}_1\tilde{\chi}^{\pm}_1$ & 5.9 pb & & 642 fb & &
$\tilde{\chi}^0_2\tilde{g}$ & 140 fb & 223 fb & 0.01 fb& 0.04 fb\\
$\tilde{\chi}^{\pm}_1\tilde{\chi}^{\pm}_2$ & 18 fb& 6 pb & 38 fb & 827 fb &
$\tilde{\chi}^0_3\tilde{g}$ & 4 fb& &0.001 fb& \\
$\tilde{\chi}^{\pm}_2\tilde{\chi}^{\pm}_2$ & 56 fb& & 147 fb & &
$\tilde{\chi}^0_4\tilde{g}$ & 8 fb& &0.02 fb&\\
\hline
$\tilde{\chi}^0_1\tilde{\chi}^0_1$ & 7 fb& &3 fb & &
$\tilde{\chi}^{\pm}_1\tilde{g}$ & 290 fb &  & 0.02 fb& \\
$\tilde{\chi}^0_1\tilde{\chi}^0_2$ & 2 fb& &2 fb& &
$\tilde{\chi}^{\pm}_2\tilde{g}$ & 20 fb& 310 fb &0.05 fb & 0.07 fb\\
\cline{6-10}
$\tilde{\chi}^0_1\tilde{\chi}^0_3$ & 6 fb& &119 fb& &
$\tilde{\chi}^{\pm}_1\tilde{\chi}^0_1$ & 227 fb&  &451 fb & \\
$\tilde{\chi}^0_1\tilde{\chi}^0_4$ & 1 fb& &$\sim 0$& &
$\tilde{\chi}^{\pm}_1\tilde{\chi}^0_2$ & 11.67 pb & & 848 fb & \\
$\tilde{\chi}^0_2\tilde{\chi}^0_2$ & 12 fb& 99 fb  &0.08 fb& 310 fb &
$\tilde{\chi}^{\pm}_1\tilde{\chi}^0_3$ & 41 fb& & 496 fb &\\
$\tilde{\chi}^0_2\tilde{\chi}^0_3$ & 18 fb& &166 fb& &
$\tilde{\chi}^{\pm}_1\tilde{\chi}^0_4$ & 17 fb& 12.2 pb & 37 fb & 2.2 pb \\
$\tilde{\chi}^0_2\tilde{\chi}^0_4$ & 2 fb& &0.2 fb& &
$\tilde{\chi}^{\pm}_2\tilde{\chi}^0_1$ & 7 fb& & 0.6 fb&\\
$\tilde{\chi}^0_3\tilde{\chi}^0_3$ & 0.01 fb& &0.09 fb& &
$\tilde{\chi}^{\pm}_2\tilde{\chi}^0_2$ & 15 fb& & 41 fb&\\
$\tilde{\chi}^0_3\tilde{\chi}^0_4$ & 51 fb &&20 fb& &
$\tilde{\chi}^{\pm}_2\tilde{\chi}^0_3$ & 100 fb& &40 fb &\\
$\tilde{\chi}^0_4\tilde{\chi}^0_4$ & 0.06 fb& &0.06 fb& &
$\tilde{\chi}^{\pm}_2\tilde{\chi}^0_4$ & 104 fb& &296 fb &\\
\end{tabular}
\caption{\label{tab:DSSxsec}Next-to-leading order cross sections for
  direct production of sparticles at the LHC for DSS1 and DSS2 as
  computed by {\tt Prospino2}~\cite{prospino}. The individual rate add
  to 81.6~pb (DSS1) and 4.3~pb (DSS2).}
\end{table}

In contrast to most supersymmetric parameter analyses for the
parameter point SPS1a~\cite{sps1a}, the analysis presented in this
paper heavily relies on rate information. Note that `rate' refers to
cross sections times branching ratios for signal plus
background. Therefore, we list all cross sections as well as all
branching ratios computed at NLO in this appendix.

Table~\ref{tab:DSSxsec} lists the next-to-leading order cross sections
for direct production of supersymmetric particles at the LHC for DSS1
and DSS2 as computed by {\tt Prospino2}~\cite{prospino}.  The combined
production rate for strongly interacting particles suffers in models
with decoupled scalars most of all because usually the quark-gluon
initiated associated squark-gluino channel dominates the
supersymmetric LHC samples. Gluino pair production is still more than
a factor two larger than light-flavor squark-antisquark production, so
we can still expect sizeable SUSY samples at the LHC.  For the DSS1
parameter point the light gluino indeed yields a relatively large
cross section for the production of gluino pairs. In DSS2, gluinos are
fairly heavy and the gluino production rate is low.

The second largest contribution to SUSY production is due to the
associated production of charginos and neutralinos with 12 pb and 2 pb
in DSS1 and DSS2, respectively. In DSS1, this channel is completely
dominated by $\tilde{\chi}^{\pm}_1\tilde{\chi}^0_2$ production while in DSS2,
equivalent contributions arise from $\tilde{\chi}^{\pm}_1\tilde{\chi}^0_{1,2,3}$ and
$\tilde{\chi}^{\pm}_2\tilde{\chi}^0_4$. This is because the neutralino mass splittings
are smaller and the gaugino/higgsino content is more degenerate in the
heavier DSS2 parameter point.  Its lower value of $\mu$ increases
$\tilde{\chi}^{\pm}_1\tilde{\chi}^0_3$ production but fails to keep up with the very
light $\tilde{\chi}^0_{1,2}$ in DSS1. The only other significant contribution
to SUSY production in the light DSS1 point comes from $\tilde{\chi}^{\pm}_1$
pairs while in DSS2 there are contributions from neutralino
pairs.

Table~\ref{tab:DSSBR} lists the higher-order branching fractions of
the supersymmetric particles involved.  Most gluino decays contain
three particles in the final state, proceeding through a very virtual
squark.  Due to its wino-like nature, the $\tilde{\chi}^0_2$ decays like a $Z$
boson plus missing energy. The larger higgsino component in the DSS2
parameter point brings in decays to charginos. Again, the light
chargino decays like a $W$ boson plus missing energy.\bigskip

\begin{table}[t]

\begin{minipage}{0.3\textwidth}  \begin{tabular}{c|c|r|r}
\multicolumn{2}{c|}{} & DSS1 & DSS2 \\
\hline
\multirow{10}{*}{$\tilde g\rightarrow$}& $\tilde{\chi}^0_1~q\bar{q}$ & 15 & 8 \\
&$\tilde{\chi}^0_2~q\bar{q}$ & 30     & 11 \\
&$\tilde{\chi}^0_3~q\bar{q}$ & $<1$   & 6  \\
&$\tilde{\chi}^0_4~q\bar{q}$ & $<1$   & 12 \\
\cline{2-4}
&$\tilde{\chi}^\pm_1~q\bar{q}$ & 55   & 24 \\
&$\tilde{\chi}^\pm_2~q\bar{q}$ & $<1$ & 28 \\
\cline{2-4}
&$\tilde{\chi}^0_1~g$ &  & $<1$ \\
&$\tilde{\chi}^0_2~g$ &  & 3 \\
&$\tilde{\chi}^0_3~g$ &  & 6 \\
&$\tilde{\chi}^0_4~g$ &  & 1 \\
\end{tabular} \end{minipage}
\begin{minipage}{0.3\textwidth} \begin{tabular}{c|c|r|c|r}
 & \multicolumn{2}{c|}{DSS1} & \multicolumn{2}{c}{DSS2} \\
\hline 
\multirow{3}{*}{$\tilde{\chi}^\pm_1\rightarrow$}&$\tilde{\chi}^0_1~q\bar{q}$ & 67 &$\tilde{\chi}^0_1~q\bar{q}$ & 66 \\ 
&$\tilde{\chi}^0_1~\ell~\nu$ & 22 & $\tilde{\chi}^0_1~\ell~\nu$ & 22 \\
&$\tilde{\chi}^0_1~\tau~\nu$ & 11 & $\tilde{\chi}^0_1~\tau~\nu$ & 11 \\
\hline
\multirow{4}{*}{$\tilde{\chi}^\pm_2\rightarrow$}&$\tilde{\chi}^\pm_1~Z$ & 32 &$\tilde{\chi}^\pm_1~Z$ & 32 \\
&$\tilde{\chi}^0_1~W$ & 8 & $\tilde{\chi}^\pm_1~h$ & 9 \\
&$\tilde{\chi}^0_2~W$ & 40 & $\tilde{\chi}^0_2~W$ & 38 \\
&$\tilde{\chi}^\pm_1~h$ & 20 & $\tilde{\chi}^0_3~W$& 19 \\
\end{tabular} \end{minipage}
\begin{minipage}{0.3\textwidth} \begin{tabular}{c|c|r|c|r}
 & \multicolumn{2}{c|}{DSS1} & \multicolumn{2}{c}{DSS2}\\
\hline 
\multirow{5}{*}{$\tilde{\chi}^0_2\rightarrow$}&$\tilde{\chi}^0_1~q\bar{q}$ & 69 &\multicolumn{2}{r}{68}  \\                                   
&$\tilde{\chi}^0_1~\ell~\ell$   & 7  & \multicolumn{2}{r}{7} \\
&$\tilde{\chi}^0_1~\tau~\tau$   & 3  & \multicolumn{2}{r}{3} \\
&$\tilde{\chi}^0_1~\nu~\bar\nu$ & 21 & \multicolumn{2}{r}{20} \\
&$\tilde{\chi}^\pm_1~q\bar{q}$  & 0  & \multicolumn{2}{r}{1} \\
\hline
\multirow{6}{*}{$\tilde{\chi}^0_3\rightarrow$}&$\tilde{\chi}^0_1~Z$ & 9 & $\tilde{\chi}^0_1~q\bar{q}$& 65 \\
&$\tilde{\chi}^0_2~Z$ & 21 &$\tilde{\chi}^0_1~\ell~\ell$ & 7 \\
\cline{2-5}
&\multirow{2}{*}{$\tilde{\chi}^\pm_1~W$} & \multirow{2}{*}{64} & $\tilde{\chi}^\pm_1~q\bar{q}$ & 3 \\
                                                  &  & & $\tilde{\chi}^\pm_1~\ell~\nu$ & 1 \\
\cline{2-5}
&$\tilde{\chi}^0_1~h$ & 2 &$\tilde{\chi}^0_1~\nu~\nu$& 20 \\
&$\tilde{\chi}^0_2~h$ & 2 &$\tilde{\chi}^0_1~\tau~\tau$& 3\\
\hline
\multirow{5}{*}{$\tilde{\chi}^0_4\rightarrow$}&$\tilde{\chi}^0_1~Z$ & 4 & \multicolumn{2}{r}{0} \\  
&$\tilde{\chi}^0_2~Z$ & 5 & \multicolumn{2}{r}{2} \\
\cline{2-5}
&$\tilde{\chi}^0_1~h$ & 6 &$\tilde{\chi}^0_3~Z$& 9 \\
\cline{2-5}
&$\tilde{\chi}^0_2~h$ & 14 & \multicolumn{2}{r}{4} \\
\cline{2-5}
&$\tilde{\chi}^\pm_1~W$ & 70  & \multicolumn{2}{r}{86} \\
\end{tabular} \end{minipage}
\caption{\label{tab:DSSBR}Branching fractions for both scenarios computed by {\tt
    SDECAY}~\cite{s_hit,sdecay}. Values rounded to the full percentage.}
\end{table}

\section{Error treatment}

As discussed in this paper, the focus on rate measurements in the case
of heavy scalars forces us to generally treat the theory uncertainties
on rate observables as possibly the dominant error at the LHC. At the
same time, some of the production rates are small, so our Markov chain
might run into parameter regions with small event numbers which have
to be treated using Poisson instead of Gaussian statistics.

Like in all SFitter analyses~\cite{sfitter,sfitter_higgs} we follow
the Rfit scheme~\cite{rfit} to combine Gaussian experimental
and flat theory errors. This scheme interprets theory errors as a lack
of knowledge on a parameter. As long as the deviation between theory
and experiment is within the theory error, this must not have any
influence on the total likelihood. Once the difference becomes larger
the (perturbative) theory is simply ruled out instead of just very,
very unlikely.  This means that we cannot simply convolute some kind of
theory likelihood distribution with a Gaussian experimental
error. Even assuming a flat theory likelihood curve would give
the difference of two one-sided error functions, \ie a peaked
likelihood. Instead, the combined log-likelihood which can also be
derived using a profile likelihood ansatz is given by
\begin{equation}
- 2 \log L = \chi^2=
\left\{\begin{array}{cr}
       0 & \text{for}~|d_i-\overline d_i|<\sigma_i^\text{th}\\
      \left(\dfrac{|d_i-\overline d_i|-\sigma_i^\text{th}}
                  {\sigma_i^\text{exp}}\right)^2 & \qquad 
             \text{for}~|d_i-\overline d_i|\geq\sigma_i^\text{th}
\end{array}\right. .
\end{equation}

For large enough event numbers the experimental error is a combination
of three different sources, all Gaussian and summed in quadrature. The
statistical error is uncorrelated between different measurements. A
first systematic error originates from the lepton energy scale and the
second from the hadronic energy scale. They are treated
separately. Each is taken as 99\% correlated between different
observables.

For smaller signal numbers systematic and statistical uncertainties
are incorporated by convoluting a Poisson probability with a Gaussian
probability, where the number of background events $N_b$ is the mean
and $\delta_b$ (systematic uncertainties) the standard
deviation~\cite{cousins_highland}. The probability that the background
fluctuates at least to the observed number of events
$N_\text{obs}=N_\text{signal}+N_\text{b}$ is
\begin{equation}
p=A \int_0^\infty db \; \text{Gauss}(N_b,\delta_b)
  \sum_{j=N_\text{obs}}^\infty\frac{e^{-b}b^j}{j!}
\end{equation}
where $A$ normalizes the integral. Then the significance of the signal
reads $Z_n=\sqrt{2} \, \text{erf}^{-1}(1-2p)$. If $N_\text{obs}$ is very
large compared to $N_b$, this significance is approximated
by~\cite{zhang_ramsden}
\begin{equation}
Z_0 = \frac{2}{\sqrt{1+\delta_b^2/N_b}}
      \left(  \sqrt{N_\text{obs}+\frac{3}{8}}
             -\sqrt{N_b+\frac{3}{8}\frac{\delta_b^2}{N_b}}
      \right).
\end{equation}


\end{document}